\documentclass[%
 reprint,
superscriptaddress,
twocolumn,
nofootinbib,
 amsmath,amssymb,
 aps,
]{revtex4-2}

\usepackage{graphicx}
\usepackage{dcolumn}
\usepackage{bm}
\usepackage{breqn}
\usepackage{xcolor}

\begin{document}

\title{Anomalous Hall effect emerging from field-induced Weyl nodes in SmAlSi}%

\author{Yuxiang Gao}
\affiliation{Department of Physics and Astronomy$,$ Rice University$,$ Houston$,$ Texas 77005$,$ USA}
\affiliation{Rice Center for Quantum Materials$,$ Rice University$,$ Houston$,$ Texas 77005$,$ USA}
\author{Shiming Lei}
\email{current affiliation: Department of Physics$,$ Hong Kong University of Science and Technology$,$ Clear Water Bay$,$ Hong Kong$,$ China}
\affiliation{Department of Physics and Astronomy$,$ Rice University$,$ Houston$,$ Texas 77005$,$ USA}
\affiliation{Rice Center for Quantum Materials$,$ Rice University$,$ Houston$,$ Texas 77005$,$ USA}%
\author{Eleanor M. Clements}
\affiliation{NIST Center for Neutron Research$,$ National Institute of Standards and Technology$,$ Gaithersburg$,$ Maryland 20899$,$ USA}
\affiliation{Materials Science and Technology Division$,$ Oak Ridge National Laboratory$,$ Oak Ridge$,$ Tennessee 37831$,$ USA}
\author{Yichen Zhang}
\affiliation{Department of Physics and Astronomy$,$ Rice University$,$ Houston$,$ Texas 77005$,$ USA}
\affiliation{Rice Center for Quantum Materials$,$ Rice University$,$ Houston$,$ Texas 77005$,$ USA}%
\author{Xue-Jian Gao}
\affiliation{Department of Physics$,$ Hong Kong University of Science and Technology$,$ Clear Water Bay$,$ Hong Kong$,$ China}%
\author{Songxue Chi}
\affiliation{Neutron Scattering Division$,$ Oak Ridge National Laboratory$,$ Oak Ridge$,$ Tennessee 37831$,$ USA}
\author{Kam Tuen Law}
\affiliation{Department of Physics$,$ Hong Kong University of Science and Technology$,$ Clear Water Bay$,$ Hong Kong$,$ China}
\author{Ming Yi}
\affiliation{Department of Physics and Astronomy$,$ Rice University$,$ Houston$,$ Texas 77005$,$ USA}
\affiliation{Rice Center for Quantum Materials$,$ Rice University$,$ Houston$,$ Texas 77005$,$ USA}
\author{Jeffrey W. Lynn}
\affiliation{NIST Center for Neutron Research$,$ National Institute of Standards and Technology$,$ Gaithersburg$,$ Maryland 20899$,$ USA}
\author{Emilia Morosan}
\email[corresponding author: E. Morosan ]{emorosan@rice.edu}
\affiliation{Department of Physics and Astronomy$,$ Rice University$,$ Houston$,$ Texas 77005$,$ USA}
\affiliation{Rice Center for Quantum Materials$,$ Rice University$,$ Houston$,$ Texas 77005$,$ USA}

\date{\today}

\begin{abstract}
The intrinsic anomalous Hall effect (AHE) has been reported in numerous ferromagnetic (FM) Weyl semimetals. However, AHE in the antiferromagnetic (AFM) or paramagnetic (PM) state of Weyl semimetals has been rarely observed experimentally. Different mechanisms have been proposed to account for the emergence of AHE from different types of magnetic order. In this paper, we propose a new model that explains the observed AHE in both the AFM and PM states of non-centrosymmetric Weyl semimetal SmAlSi. The newly-proposed mechanism is based on magnetic field-induced Weyl nodes evolution, which qualitatively explains the temperature dependence of the anomalous Hall conductivity (AHC), which displays unconventional power law behavior in both the AFM and PM states of SmAlSi.  
\end{abstract}

\maketitle


Magnetic Weyl semimetals have garnered a lot of interest in recent years because of the interplay between electronic correlations and topology \cite{Fujishiro2016, Yang2017, Liu2019, Puphal2020, Dzsaber2021, Gaudet2021}. Magnetism drastically increases the number of possible symmetries, from 230 crystalline space groups to 1651 magnetic space groups if we only consider the commensurate order, allowing for the tunability of topology through the magnetic structure. In magnetic Weyl semimetals, Weyl fermions can mediate magnetic order through nesting between different Weyl pockets \cite{ Gaudet2021, Lifshitz2022, Watanabe2018, Jennifer2019, Wang2022}. Furthermore, the Weyl nodes of different chirality in magnetic Weyl semimetals bring about a broad range of magneto-transport and spectroscopic properties \cite{Hirschberger2016, Shekar2018, Sanchez2020, Morali2019, Li2020, Liu2016}. Among these, the anomalous Hall effect (AHE) in Weyl semimetals stands out, especially in contrast to the AHE in normal ferromagnets, \textit{e.g.,} iron \cite{Feng1975, Taguchi2004, Chun2007, Miyasato2007, Shiomi2009}. In the latter, the spin-orbit interaction and the net magnetization M are essential to establishing a non-zero AHE, where the anomalous Hall conductivity (AHC) is found to be proportional to M in FM materials  \cite{Nagaosa2010}. By contrast, the Weyl nodes behave as effective magnetic monopoles, generating strong Berry curvature which, in turn, acts like an effective magnetic field in momentum space. In this scenario, the FM order is no longer required for a Weyl semimetal to host non-zero AHE \cite{Armitage2018, Jin2019, Tian2018}. 

To date, AHE has been reported in a few magnetic Weyl semimetals \cite{Zeng2006, Nayak2016, Nakatsuji2016, Suzuki2016, Lee2007, Wang2016, Guin2020, Kim2018, Singh2021}, associated with different mechanisms. The AHE shows linear dependence on the magnetization M in some FM compounds ($\mathrm{Fe_3Sn_2}$, MnSi, PrAlGe, $\mathrm{Mn_5Ge_3}$, $\mathrm{Co_3Sn_2S_2}$ and $\mathrm{Fe_3GeTe_2}$)\cite{Zeng2006, Lee2007, Wang2016, Guin2020, Kim2018, Singh2021}, but not in AFM and non-magnetic (NM) materials ($\mathrm{ZrTe_5}$, (Nd, Gd)PtBi, $\mathrm{Mn_3(Ge,Sn)}$ and $\mathrm{KV_3Sb_5}$) \cite{Nayak2016, Nakatsuji2016, Suzuki2016, Tian2018, Yang2020}.  These AFM or NM Weyl semimetals do not host Weyl nodes without further breaking the time-reversal symmetry. Different AHE mechanisms have recently been proposed to explain the AHE in the absence of FM order \cite{Fujishiro2016, Burkov2014, Chen2014}. Though supported theoretically, the AHE has not been found in materials that host Weyl nodes without time-reversal symmetry breaking. This precludes further development of a theoretical understanding of the AHE mechanisms in these materials. 

Here, we report large AHE in the AFM and PM state of SmAlSi, which is a member of the non-centrosymmetric class of compounds \textit{R}Al(Si,Ge) (\textit{R} = La, Ce, Pr, Nd, Sm) \cite{Gaudet2021,Puphal2020,Destraz2020,Chang2018,Xu2017,Sanchez2020,Suzuki2019,Xu2017,Yao2022}. Moreover, we propose a new mechanism to account for the AHE in non-FM materials without a center of symmetry. This mechanism is based on the displacement of the Weyl nodes to points with different momenta upon the application of a magnetic field, which results in regions with non-zero Chern numbers in the first Brillouin zone. The Weyl points and Kramers nodal lines (KNLs) in the PM state of SmAlSi have been investigated through density functional theory (DFT) calculations, angle-resolved photoemission spectroscopy (ARPES) and angle-dependent quantum oscillation (QO) measurements \cite{Zhang2022}. In addition to the previous report of the topological Hall effect (THE) in the intermediate-field A phase \cite{Yao2022}, here we report large AHE in the AFM ground state \textit{and} in the PM state of SmAlSi up to 100 K. This is distinct from the AHE in CeAlSi, CeAlGe, PrAlSi and PrAlGe \cite{Puphal2020, Destraz2020, Yang2021, Meng2020}, which was discovered only in the FM state, and was found to be proportional to M. This is the first time that a non-zero AHE is reported in a non-centrosymmetric Weyl semimetal outside of a FM state. 



\begin{figure}
\includegraphics[width=0.5\textwidth]{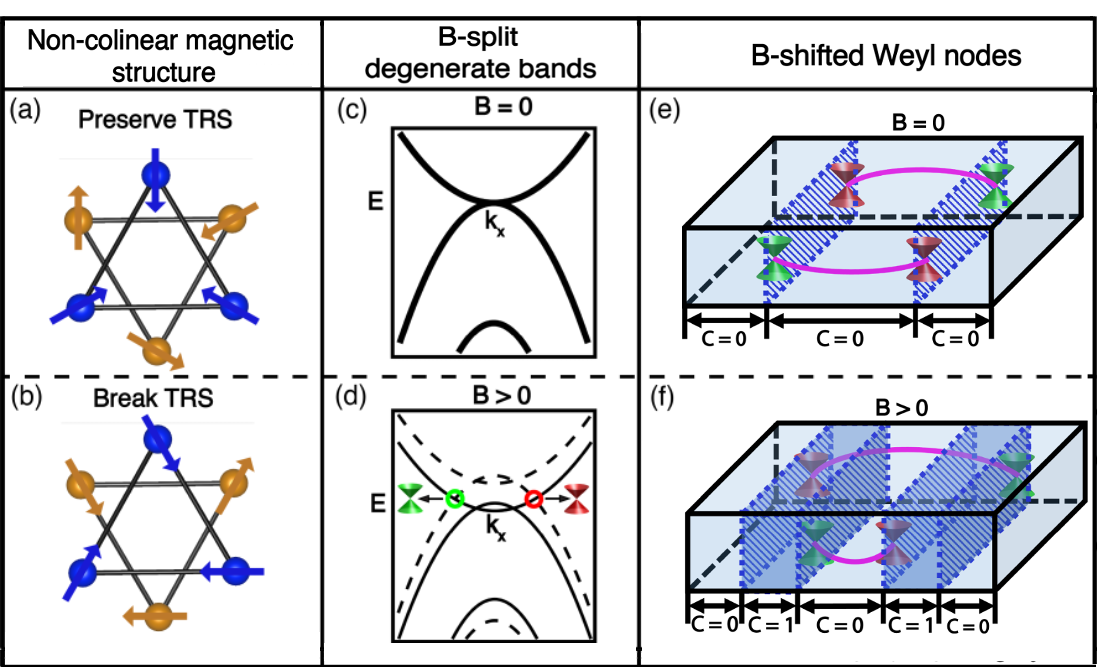} 
\caption{\label{fig:1} Proposed AHE mechanisms in $\mathrm{Mn_3Sn}$, GdPtBi and SmAlSi, based on (a-b) breaking \textit{T} in non-collinear magnetic structures, (c-d) field-split degenerate bands, or (e-f) field-shifted Weyl nodes. (a-b) Comparison of two possible AFM magnetic structures in $\mathrm{Mn_3Sn}$ which forbid any finite AHE (a) or allow finite AHE (b). (c-d) Comparison of the GdPtBi band structure without (c) and with (d) applied magnetic field. The magnetic field lifts the band degeneracy and induces Weyl points. (e-f) Comparison of SmAlSi band structure without (e) and with (f) applied magnetic field. The Weyl nodes with different chiralities are illustrated in red and green, with Fermi arcs connecting them (magenta curves). The magnetic field changes the position of the Weyl nodes and induces a region with non-zero Chern number (c) in the Brillouin zone.}
\end{figure}



Existing proposals for mechanisms responsible for AHE in AFM materials prove inadequate for the non-centrosymmetric SmAlSi. First principles calculations \cite{Chen2014} demonstrate the possibility of non-zero AHE in AFM materials with non-colinear magnetic structures. The non-colinear magnetic structure can spontaneously break the time-reversal symmetry ({\it T}). In the AFM kagomé metal $\mathrm{Mn_3Ir}$, breaking {\it T} results in non-zero Berry curvature and non-zero AHC. In the isostructural $\mathrm{Mn_3Sn}$ and $\mathrm{Mn_3Ge}$, the Mn magnetic sublattice determines if {\it T} is broken, which, in turn, determines if there is non-zero AHE, as illustrated in Fig.~\ref{fig:1}(a-b). The magnetic structure in Fig.~\ref{fig:1}(a) is a bipartite lattice (antiparallel magnetic moment pairs) and preserves {\it T}, while the magnetic structure in Fig.~\ref{fig:1}(b) carries a winding chirality and break {\it T}. If the Weyl points are close to the Fermi energy E$_F$, the amplitude of the AHE is enhanced compared to cases where the Weyl points are away from E$_F$ \cite{Kubler2014}.  Overall, the broken {\it T} and the Weyl points created by the broken symmetry could explain the observation of the AHE in the ordered state of $\mathrm{Mn_3Sn}$ and $\mathrm{Mn_3Ge}$ \cite{Nayak2016, Nakatsuji2016}. 

In stark contrast to $\mathrm{Mn_3Sn}$ and $\mathrm{Mn_3Ge}$, neutron scattering and $\mu$SR measurements show that the magnetic order in GdPtBi is collinear AFM type below $\mathrm{T_N}$ = 9.2~K \cite{Suzuki2016,Shekar2018}. Despite the lack of inversion center, the electronic structure of GdPtBi in zero field does not feature any Weyl nodes  \cite{Suzuki2016,Shekar2018}. However, AHE is observed up to 50 K, which is significantly larger than $\mathrm{T_N}$. Despite these differences, the GdPtBi hosts non-trivial topology, similarly to $\mathrm{Mn_3Sn}$ and $\mathrm{Mn_3Ge}$. By breaking {\it T} via an external magnetic field, the degeneracy of the electronic bands is lifted through the exchange interaction between the Gd ions \cite{Shekar2018,Hirschberger2016}. As illustrated in Fig.~\ref{fig:1}(c-d), the combination of Zeeman effect and magnetic exchange interaction splits the electronic bands, leading to the formation of Weyl nodes in GdPtBi \cite{Shekar2018,Hirschberger2016}. 
This mechanism is responsible for the observed AHE as well as the anomalous Nernst effect, negative longitudinal magnetoresistance, and planar Hall effect in GdPtBi \cite{Suzuki2016,Shekar2018,Hirschberger2016,Zhu2023,Liang2018}. 

Weyl nodes can occur in non-centrosymmtric materials even in the absence of an applied magnetic field, as is the case in \textit{R}Al(Si,Ge) \cite{Chang2018,Xu2017}. In {\it I}-breaking Weyl semimetals, if {\it T} is preserved, Weyl nodes with different chiralities appear at opposite momenta, and all regions in the first Brillouin zone have zero Chern number, such that the AHE is expected to be zero [Fig.~\ref{fig:1}(e)]. FM order spontaneously breaks{\it T} \cite{Chang2018}, and the Weyl nodes are effectively shifted by the exchange interaction, resulting in non-zero integration of the Berry curvature that leads to non-zero AHE. Here we extend this scenario to materials without FM order. An external magnetic field could break {\it T}, and the magnetic exchange interaction could act as FM exchange. As a result, there are regions with non-zero Chern number in the first Brillouin zone [Fig.~\ref{fig:1}(f)], leading to a field-induced non-zero AHE. It is worth noting that the Zeeman energy is typically significantly smaller than the magnetic exchange energy. Therefore, the AHE resulting from this mechanism is expected to be substantially smaller in NM materials compared to magnetic materials. In the magnetic system SmAlSi, we expect a nonzero AHE even in the PM state, particularly within the temperature range just above the Ne\'el temperature $T_N$.   


Our earlier work has shown that SmAlSi orders antiferromagnetically below $T_N$ = 11.3 K, and undergoes a second magnetic transition at $T\mathrm{_1}$ = 4.1 K \cite{Zhang2022}. Recent neutron scattering measurements by Yao \textit{et al.} revealed the incommensurate AFM nature of the magnetic order below $T_N$ \cite{Yao2022}. Magnetic Bragg peaks were found at \textbf{Q} = (1/3-$\delta$, 1/3-$\delta$, 4) and (1/3-$\delta$, 1/3-$\delta$, 8), which could be indexed by a propagation vector \textbf{k} = (1/3-$\delta$, 1/3-$\delta$, 0) with $\delta$ = 0.007(7). We also carried out diffraction measurements and confirmed the basic magnetic order wave vector is (1/3, 1/3, 0) with an ordering temperature of $T_N$ = 11.2(2) K [Fig. S1(a,b)], where the uncertainty represents one standard deviation.

 \begin{figure}
\includegraphics[width=0.45\textwidth]{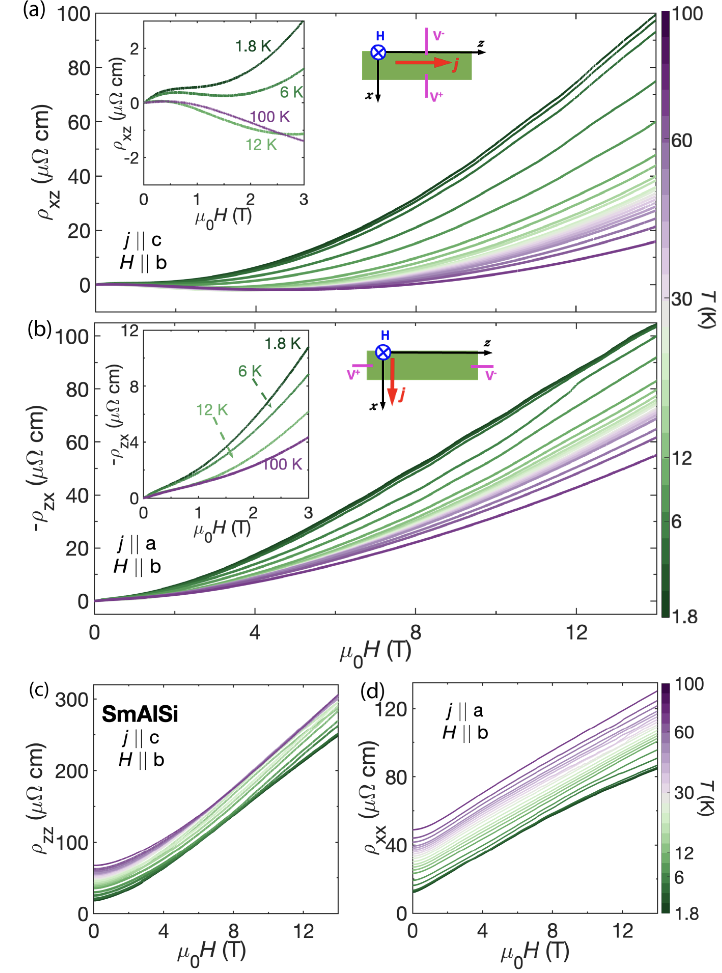} 
\caption{\label{fig:3} Transverse resistivity ($\rho_{xz}$ and $\rho_{zx}$)  longitudinal resistivity ($\rho_{zz}$ and $\rho_{xx}$ of SmAlSi).  (a, b) Field-dependent Hall resistivity $\rho_{xz}$ and $\rho_{zx}$, with current $j \parallel c$ and magnetic field $H \parallel b$ (a) and current $j \parallel a$ and magnetic field $H \parallel b$ (b), respectively. The schematics of the corresponding Hall resistivity are shown in the inset. (c,d) Field-dependent resistivity $\rho_{zz}$ and $\rho_{xx}$ as a function of magnetic field.
}
\end{figure}

Magnetotransport measurements with current along the \textit{c}/\textit{a} axis and magnetic field parallel to the \textit{b} axis revealed large anomalous Hall contributions to the resistivity and conductivity in both the AFM and PM states. The non-linear field-dependent Hall resistivity components $\rho_{xz}$ and $\rho_{zx}$ [Fig. \ref{fig:3}(a,b)] indicate the coexistence of electron and hole carriers, consistent with previous studies \cite{Gaudet2021, Destraz2020, Cao2022}, although these measurements were performed for a different current-field configuration ($j~\parallel~a$, $H~\parallel~c$). Surprisingly, the Hall resistivity $\rho_{xz}$ and $\rho_{zx}$ shows an unusual plateau between 0 and 2 T [Fig. \ref{fig:3}(a,b), inset], which was not observed before, likely due to the different measurement configuration \cite{Gaudet2021, Destraz2020, Cao2022}. 

\begin{figure}[h]
\includegraphics[width=0.45\textwidth]{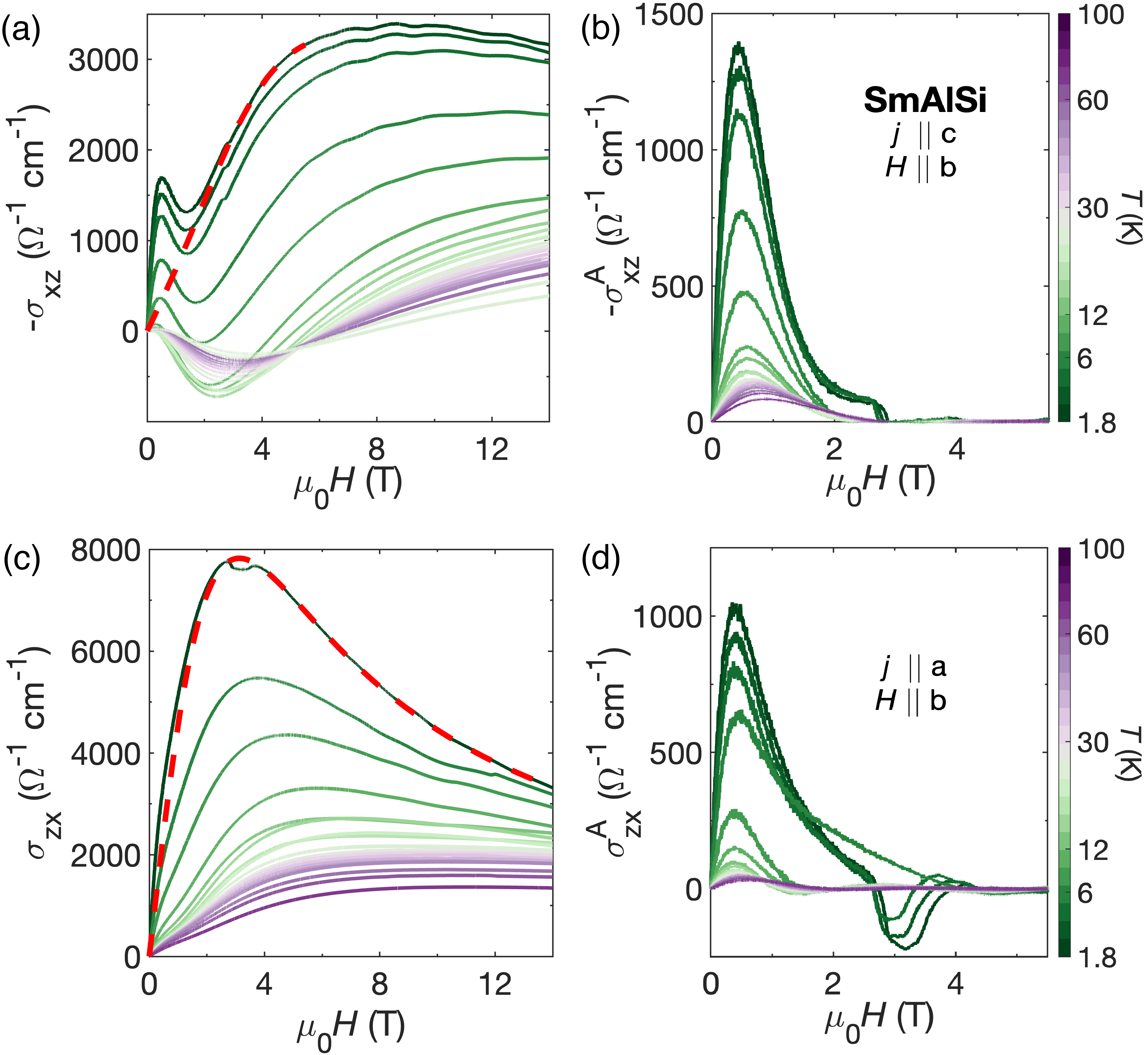} 
\caption{\label{fig:5} 
The AHC of SmAlSi. (a,c) Field-dependent Hall conductivity $\sigma_{xz}$ and $\sigma_{zx}$ of SmAlSi. The red dashed lines are the two-band fit ordinary Hall contribution to the Hall conductivity at 1.8 K. (b,d) AHC $\sigma^A_{xz}$ and  $\sigma^A_{zx}$ as a function of the magnetic field at different temperatures T = 1.8 K to 100 K.
}
\end{figure}

Considering the additive Hall conductivity contributions \cite{Nagaosa2010}, we use conductivity (rather than resistivity) analysis to probe the existence of the anomalous Hall effect. Since \textit{c} axis is not equivalent to the \textit{a} or \textit{b} axis, the conductivity tensor and the resistivity tensor are connected through 
\begin{equation}\label{eq0}
    \begin{bmatrix}
       \sigma_{xx} & \sigma_{xz}\\
       \sigma_{zx} & \sigma_{zz}
    \end{bmatrix} = 
    \frac{1}{\rho_{xx}*\rho_{zz}-\rho_{xz}*\rho_{zx}}
    \begin{bmatrix}
       \rho_{xx} & -\rho_{xz}\\
       -\rho_{zx} & \rho_{zz}
     \end{bmatrix} 
\end{equation}
Fig. \ref{fig:3}(a,b) shows the measured resistivity data for the relevant current (j) and field (H) orientations, which are used in eq. \ref{eq0} to estimate the conductivity components $\sigma_{xz}$ and $\sigma_{zx}$ (Fig. \ref{fig:5}(a,c)). The unusual plateau between 0 and 2 T in $\rho_{xz}$ develops into an additional peak in $\sigma_{xz}$ and a broad maximum in $\sigma_{zx}$. 

To rule out a simple multi-band explanation for the observed magnetotransport, we performed two-band model fits to capture the low-field anomaly at 1.8 K. An example of such a fitting is shown in Fig. S5 (dashed lines) for {\it n}$_1$ = 1.08*$\mathrm{10^{20} cm^{-3}}$, $\mu_1$ = 606 $\mathrm{cm^{2}V^{-1}s^{-1}}$ and {\it n}$_2$ = 5.36*$\mathrm{10^{17} cm^{-3}}$, $\mu_2$ = 30436 $\mathrm{cm^{2}V^{-1}s^{-1}}$. However, the best fits do not capture the Hall conductivity well for the entire field range. In addition, the conductivity from the fit $\sigma^{fit}_{xx}(H=0)$ = ${n_1e\mu_1}+{n_2e\mu_2}$ = 1.31*$\mathrm{10^{4} \Omega^{-1}cm^{-1}}$ is significantly smaller than the experimental value $\sigma_{xx}(H=0)$ = 8.18*$\mathrm{10^{4} \Omega^{-1}cm^{-1}}$. These observations suggest that a simple multi-band model without the inclusion of an anomalous term is not adequate to account for the Hall conductivity. 

\begin{figure}
\includegraphics[width=0.45\textwidth]{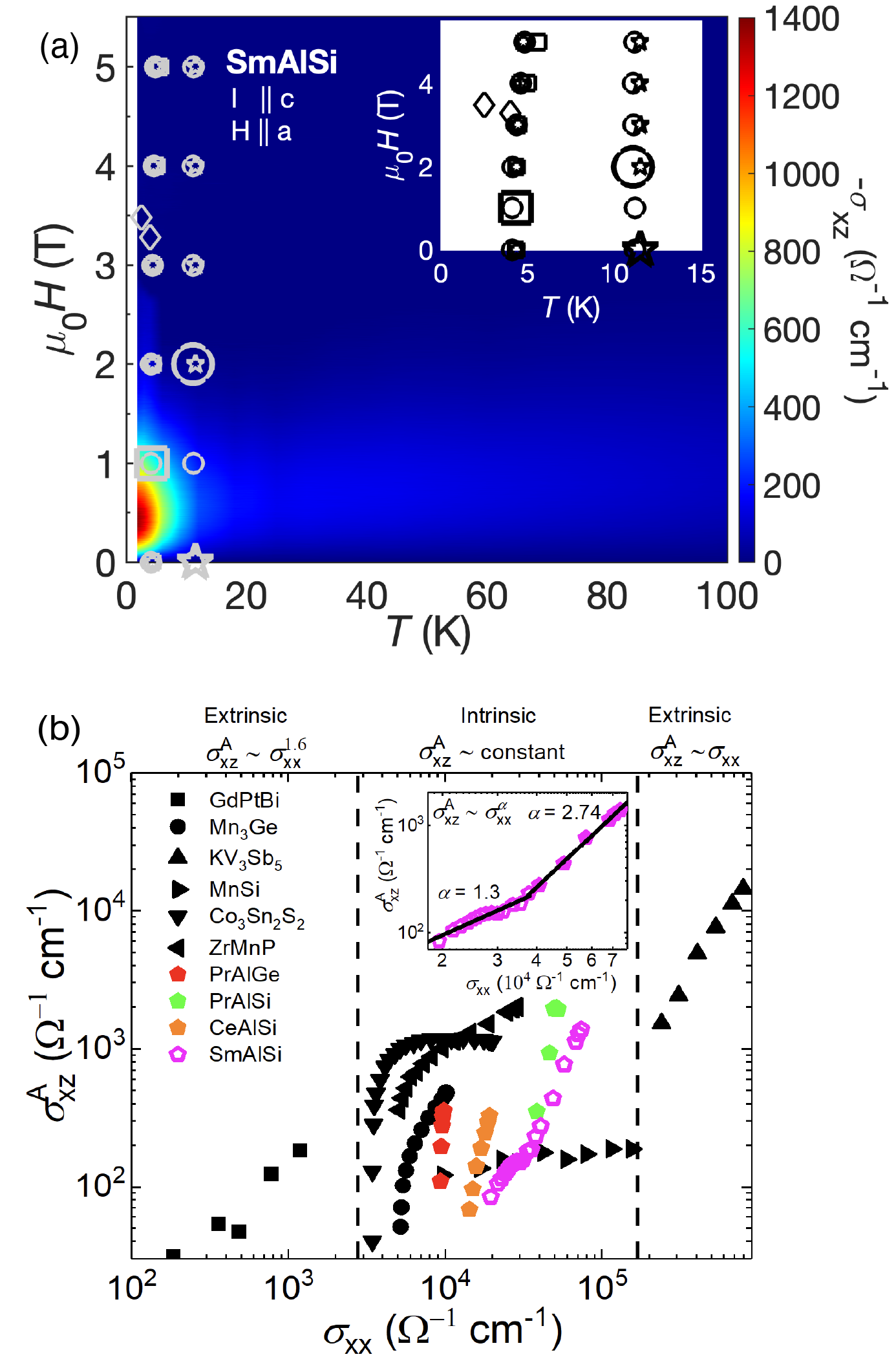} 
\caption{\label{fig:4} (a) Contour map of $\sigma^A_{xz}$ for $\mu_0$H = 0 - 14 T and T = 1.8 - 100 K. The symbols are the phase boundaries determined from Figs. S1-3. Inset: a zoomed-in view of the phase diagram at low temperatures. The phase boundaries are determined from the peaks in Cp(T) (stars), $\mathrm{d(M(T)/H)/dT}$ (squares), {d$\rho$(T)/dT} (circles), $\mathrm{dM(H)/dH}$ (diamonds). (b) $\sigma^A_{xz}$ \textit{vs.} $\sigma^A_{xx}$ for SmAlSi (magenta pentagons), other RAl(Si,Ge) (red, green, and orange pentagons) \cite{Destraz2020,Meng2020,Yang2021}, and other topological semimetals (black symbols) \cite{Suzuki2016,Nayak2016,Yang2020,Lee2007,Liu2016,Singh2021}.
}
\end{figure}

For magnetic materials, the Hall conductivity sums up two contributions, if the topological spin texture is not considered: the ordinary Hall effect $\sigma^{N}$, and the AHE $\sigma^{A}$ \cite{Nagaosa2010}:
\begin{equation}\label{eq1}
   \sigma_{xz}=\sigma^{N}_{xz}+\sigma^{A}_{xz},
\end{equation}
where the AHE is proportional to the magnetization M 
$\sigma^{A}_{xz}=S_HM$. Assuming the linear field dependence of the Hall resistivity, and a small Hall angle ($\rho_{xz}~<<~\rho_{zz}$), the measured Hall resistivity $\rho_{xz}$ can be written as:
\begin{equation}\label{eq2}
   \rho_{xz}~=~R_0H~+~S_H~\rho_{zz}^2M.
\end{equation}
While this formula has been successfully applied to analyze the AHE in many FMs, it fails to describe high-mobility multi-band systems. In GdPtBi for example, the field-dependent normal Hall effect is nonlinear due to the multi-band transprt\cite{Suzuki2016,Shekar2018}. Instead, the AHE is estimated by subtracting the scaled Hall resistivity measured far above the AFM ordering temperature ($T_N~=~9.2$ K). Such analysis is valid assuming the carrier density and mobility only slightly vary below this temperature. This may not be the case for SmAlSi. To capture the nonlinear field dependence of the Hall effect, we performed a two-band fit to the high field Hall conductivity data $\sigma_{xz}$ and $\sigma_{zx}$. Example fits are shown as dashed lines in Fig. \ref{fig:5}(a,c). The difference between the data and model fits $-\sigma^A_{xz}$ [Fig.~\ref{fig:5}(b,d)] reflects the AHE contribution. A contour plot of this transport response in the \textit{H} - \textit{T} phase diagram is shown in Fig. \ref{fig:4}(a). Such AHE map bears a striking resemblance to that established for GdPtBi \cite{Suzuki2016,Shekar2018}: both compounds display a peak at low fields, after subtracting the Hall conductivity fit to the high field data. Moreover, the peak persists far above the AFM transition temperature, although the amplitude is reduced with increasing T. In GdPtBi, the AHE persists up to $\sim~50$ K  when $T_N~=~9.2$ K \cite{Suzuki2016,Shekar2018}. By comparison, it persists up to temperatures twice as high {\it T} $\sim~100$ K in SmAlSi, even when the ordering temperature is comparable $T_N~=~11.3$ K. At 1.8 K, $\sigma^{A}_{xz}$ and $\sigma^{A}_{zx}$ reach 1380 and 1030 \textit{$\Omega^{-1} cm^{-1}$}, about one order of magnitude larger than the values in GdPtBi ($\sim$ 30 - 200 \textit{$\Omega^{-1} cm^{-1}$}). The resulting tangent of the anomalous Hall angle, $tan \theta^{A}_{xz} = \sigma^{A}_{xz}/\sigma_{xx}$ is 0.017 and $tan \theta^{A}_{zx} = \sigma^{A}_{zx}/\sigma_{zz}$ is 0.02, which is about 1/8 to 1/10 of that observed in GdPtBi \cite{Suzuki2016,Shekar2018}, but comparable to that of CeAlSi (0.02) or PrAlGe (0.025) \cite{Destraz2020,Cheng2023}, although the latter systems are both ferromagnets. 

Compared with results from \cite{Yao2022} where the Hall anomaly is reported in the finite-field A phase, the Hall anomaly from our Hall conductivity data is found in a much larger temperature range and a distinct magnetic field range. In addition to a different measurement geometry (different current and field direction), the measurements are done on two different samples with better crystal quality (as demonstrated by heat capacity \cite{Zhang2022} and quantum oscillations data (SI Figs. 7-8)). In addition, the differences in the quantum oscillation frequencies and Hall resistivity indicate that the Fermi energy in our samples is lower than in the samples from \cite{Yao2022}.

The contour map of the AHE contribution in SmAlSi [Fig. \ref{fig:4}(a)] shows a striking resemblance to that in GdPtBi \cite{Suzuki2016}. Weyl nodes are expected to play an important role in both systems. Nevertheless, the mechanisms for the formation of the Weyl nodes are quite different. The electronic structure of GdPtBi in zero field does not feature any Weyl nodes. Under a finite field, Weyl nodes form due to the field-induced Zeeman splitting combined with the effect of AFM exchange field. Recently, an AHE with an amplitude much larger than GdPtBi was discovered in TbPdBi \cite{Zhu2023}. The study proposed a crossover from a Berry phase cancellation in the AFM/PM state to a half-topological state with large non-zero Berry curvature (hosted by anti-crossing points) in the FM state based on the band structure calculation \cite{Zhu2023}. The Berry curvature in the reciprocal space is responsible for the AHE in both compounds. The formation of the anti-crossing points in TbPdBi is similar to the Weyl points in GdPtBi: exchange field of the RE ions in the corresponding compounds. In contrast, the Weyl nodes in SmAlSi exist even at zero field above $T_N$, due to {\it I} breaking and SOC.  

The AHE in SmAlSi is also significantly different from that in $\mathrm{Mn_3(Ge,Sn)}$. In the latter, AHE is observed below the ordering temperature while in the former, it persists well above $T_N$. At H = 0, the AHE is zero in SmAlSi and non-zero in $\mathrm{Mn_3(Ge,Sn)}$. These differences highlight the importance of the magnetic order in $\mathrm{Mn_3(Ge,Sn)}$ to induce the AHE, which spontaneously breaks {\it T}, while in SmAlSi, it is the magnetic field that breaks {\it T}. 

It is necessary to discuss possible Fermi surface reconstruction under a magnetic field in the presence of AHE, as discovered in $\mathrm{YbMnBi_2}$ \cite{Pan2022}. In $\mathrm{YbMnBi_2}$, it has been proposed that the canted Mn moments could change the shape and position of the bands dramatically, such that the energy of the Weyl points relative to the Fermi energy can be tuned, and, in turn, tune the intrinsic anomalous Hall and Nerst effects \cite{Pan2022}. It should be noted that the magnetic exchange energy in $\mathrm{YbMnBi_2}$ is significantly higher than in SmAlSi: $\mu^{Mn}_{eff} \sim 4~\mu_B$ with $T_N$ = 290 K, and $\mu^{Sm}_{eff} \sim 0.85~\mu_B$ with $T_N$ = 11.3 K, respectively. The exchange energy from the canted $\mathrm{Sm^{3+}}$ ions should be too small to significantly impact the Fermi surface topology as in $\mathrm{YbMnBi_2}$. In addition, the quantum oscillation frequencies [Fig. S8] do not change across the transitions. Therefore, neither temperature nor field are likely to induce Fermi surface reconstruction in SmAlSi.

We benchmark the AHC in SmAlSi against other established topological semimetals, as indicated by the $\sigma^A_{xz}$ dependence on $\sigma_{xx}$ [Fig.~\ref{fig:4}(b)]. The conductivity of SmAlSi falls in the intrinsic regime, where the inter-band-coherence-driven intrinsic AHC is independent of the electron scattering mean free time, and should be a constant: $\sigma^A_{xz}~\sim$ constant. The amplitude of the AHC is comparable to that of other materials with intrinsic AHE. 

The extrinsic skew scattering may contribute to the AHE in SmAlSi as well. In $\mathrm{YbMnBi_2}$, the strong anisotropy of the Fermi surface leads to large anisotropy in the conductivity ($\sigma_{bb}$/$\sigma_{cc}$ $\sim$ 35). With H parallel to a, The anomalous Hall and Nerst effects are dramatically enhanced when the current or temperature gradient is parallel to b, as expected for the extrinsic mechanism. In SmAlSi, the conductivity anisotropy is small ($\sigma_{zz}$/$\sigma_{xx}$ $\sim$ 1.5), with similar anomalous Hall conductivity amplitude $\sigma^A_{xz} \sim 1380~\Omega^{-1} cm^{-1}$ and $\sigma^A_{zx} \sim 1030~\Omega^{-1} cm^{-1}$.

Apart from the AHE, a non-coplanar spin texture may yield spin chirality, generating Hall anomaly that is usually referred to as THE \cite{Tokura2021}. We will elaborate on this scenario in the AFM and PM phases of SmAlSi. 

Based on the current understanding, it is unclear if any magnetic phases have spin chirality in SmAlSi. Due to the large Sm neutron absorption cross section, neutron scattering experiments from \cite{Yao2022} and this work only show the basic magnetic order wave vector as (1/3, 1/3, 0) with the magnetic structure undetermined, while the large standard deviation precludes a solid proof of incommensurate order. Moreover, the multi-q feature is missing in the neutron scattering results on SmAlSi, which has been seen as key evidence for non-coplanar spin texture. It is unclear if the magnetic structure establishes spin chirality in SmAlSi as host of skyrmionic correlation. 

Skyrmionic correlation could also persist in a small temperature range outside the skyrmion phase, resulting in a vestigial skyrmion lattice phase, as has been shown in MnSi recently \cite{Kindervater2019}. The vestigial phase is limited to a small range close to the skyrmion phase in MnSi, while the Hall anomaly persists in a much larger range well above the ordering temperature. Therefore, it is unlikely that the Hall anomaly in SmAlSi is related to the vestigial skyrmion lattice. 

We now turn to the discussion of thermal fluctuations, which, in addition to long-range magnetic order,  have been shown to yield finite spin chirality \cite{Wang2019,Kolincio2023}. A competition of thermal fluctuations and Zeeman energy or Dzyaloshinskii–Moriya (DM) interaction has been shown to induce spin chirality in $\mathrm{Gd_3Ru_4Al_{12}}$, and in thin film of $\mathrm{SrRuO_3}$ and V-doped $\mathrm{Bi_2Se_3}$, respectively \cite{Wang2019,Kolincio2023}. However, as proposed in \cite{Kolincio2023}, the fluctuation-driven effect is promoted only when the spin chirality between neighboring plaquettes is not canceled. For lattice motifs with equivalent polygons as in SmAlSi, the fluctuation-driven chirality is expected to cancel out. 

It would therefore appear that none of the aforementioned mechanisms are adequate to describe the magnetotransport in SmAlSi. It will be helpful to further explore both the intrinsic and extrinsic contributions to the AHE in SmAlSi, as well as the magnetic structures and magnetic interactions through different theoretical and experimental approaches. The AHE in SmAlSi [Fig.~\ref{fig:4}(b)] is not a constant as expected for intrinsic deflection \cite{Nagaosa2010}. Such behavior is similar to the AHE found in the FMs CeAl(Si,Ge) and PrAl(Si,Ge) \cite{Puphal2020, Destraz2020, Yang2021, Meng2020}, despite the fact that the AHE is missing in their PM state. Here we provide a qualitative explanation of this dependence, based on a new mechanism. 


In our proposed mechanism, the Weyl nodes evolution under an external magnetic field can generate non-zero AHE in non-centrosymmetric systems. Moreover, the AHC has been shown to be proportional to the momentum space separation between Weyl nodes of different chiralities \cite{Burkov2014}. In SmAlSi, such separation is directly related to the size of the magnetic interaction between the canted $\mathrm{Sm^{3+}}$ magnetic moments. A larger moment will induce a larger shift in the electronic bands. The size of the canted moment is reflected in the susceptibility [Fig. S3(a)] or the isothermal magnetization [Fig. S3(c)]. The measurements of both quantities show a negative dependence on temperature, which is consistent with the change in the AHE.

Moreover, the $\sigma^A_{xz}$ of SmAlSi exhibits two regions that could be differentiated by distinct power-law coefficients, as illustrated by the linear fits in the inset of Fig.~\ref{fig:4}(b). These two regions coincide with the PM phase (full symbols) and the AFM phase (open symbols). To quantitatively explain this unconventional power-law behavior, extensive theory and experimental work are required to study the AHE dependence under this new mechanism.

We discovered large AHE in both the AFM and PM states of SmAlSi, which is the first observation of AHE in the non-FM state of non-centrosymmetric Weyl semimetals. We propose a new AHE mechanism in non-centrosymmetric Weyl semimetals, based on an external magnetic field breaking \textit{T}, and shifting the position of the Weyl nodes. Consequently, these Weyl nodes create regions with non-zero Chern number and generate AHE. Our work motivates further studies into the quantitative understanding of the AHE in SmAlSi and other AFM materials. 

This work was primarily supported by the Department of Defense, Air Force Office of Scientific Research under Grant No. FA9550-21-1-0343. S.L. and E.M. acknowledge partial support from the the Robert A. Welch Foundation grant C-2114.  K.T.L. acknowledges the support of HKRGC through RFS2021-6S03, C6025-19G, AoE/P-701/20, 16307622, 16310520 and 16310219. A portion of this research used resources at the High Flux Isotope Reactor, a DOE Office of Science User Facility operated by the Oak Ridge National Laboratory. The work of EMC (neutron scattering data collection and analysis) was supported by the U.S. Department of Energy (DOE), Office of Science, Basic Energy Sciences (BES), Materials Sciences and Engineering Division. The identification of any commercial product or trade name does not imply endorsement or recommendation by the National Institute of Standards and Technology.

             \end{document}



\title{Supplementary information: Anomalous Hall effect emerging from field-induced Weyl nodes in SmAlSi}

\author{Yuxiang Gao}
\affiliation{Department of Physics and Astronomy$,$ Rice University$,$ Houston$,$ Texas 77005$,$ USA}
\affiliation{Rice Center for Quantum Materials$,$ Rice University$,$ Houston$,$ Texas 77005$,$ USA}
\author{Shiming Lei}
\email{Current affiliation: Department of Physics$,$ Hong Kong University of Science and Technology$,$ Clear Water Bay$,$ Hong Kong$,$ China}
\affiliation{Department of Physics and Astronomy$,$ Rice University$,$ Houston$,$ Texas 77005$,$ USA}
\affiliation{Rice Center for Quantum Materials (RCQM) and Smalley-Curl Institute (SCI)$,$ Rice University$,$ Houston$,$ Texas 77005$,$ USA}%
\author{Eleanor M. Clements}
\affiliation{NIST Center for Neutron Research$,$ National Institute of Standards and Technology$,$ Gaithersburg$,$ Maryland 20899$,$ USA}
\affiliation{Materials Science and Technology Division$,$ Oak Ridge National Laboratory$,$ Oak Ridge$,$ Tennessee 37831$,$ USA}
\author{Yichen Zhang}
\affiliation{Department of Physics and Astronomy$,$ Rice University$,$ Houston$,$ Texas 77005$,$ USA}%
\affiliation{Rice Center for Quantum Materials$,$ Rice University$,$ Houston$,$ Texas 77005$,$ USA}
\author{Xue-Jian Gao}
\affiliation{Department of Physics$,$ Hong Kong University of Science and Technology$,$ Clear Water Bay$,$ Hong Kong$,$ China}%
\author{Songxue Chi}
\affiliation{Neutron Scattering Division$,$ Oak Ridge National Laboratory$,$ Oak Ridge$,$ Tennessee 37831$,$ USA}
\author{Kam Tuen Law}
\affiliation{Department of Physics$,$ Hong Kong University of Science and Technology$,$ Clear Water Bay$,$ Hong Kong$,$ China}%
\author{Ming Yi}
\affiliation{Department of Physics and Astronomy$,$ Rice University$,$ Houston$,$ Texas 77005$,$ USA}
\affiliation{Rice Center for Quantum Materials$,$ Rice University$,$ Houston$,$ Texas 77005$,$ USA}
\author{Jeffrey W. Lynn}
\affiliation{NIST Center for Neutron Research$,$ National Institute of Standards and Technology$,$ Gaithersburg$,$ Maryland 20899$,$ USA}
\author{Emilia Morosan}
\email[corresponding author: E. Morosan ]{emorosan@rice.edu}
\affiliation{Department of Physics and Astronomy$,$ Rice University$,$ Houston$,$ Texas 77005$,$ USA}
\affiliation{Rice Center for Quantum Materials$,$ Rice University$,$ Houston$,$ Texas 77005$,$ USA}

\date{\today}
             
\maketitle

\section{Neutron scattering}
\begin{figure*}[h]
\includegraphics[width=0.9\textwidth]{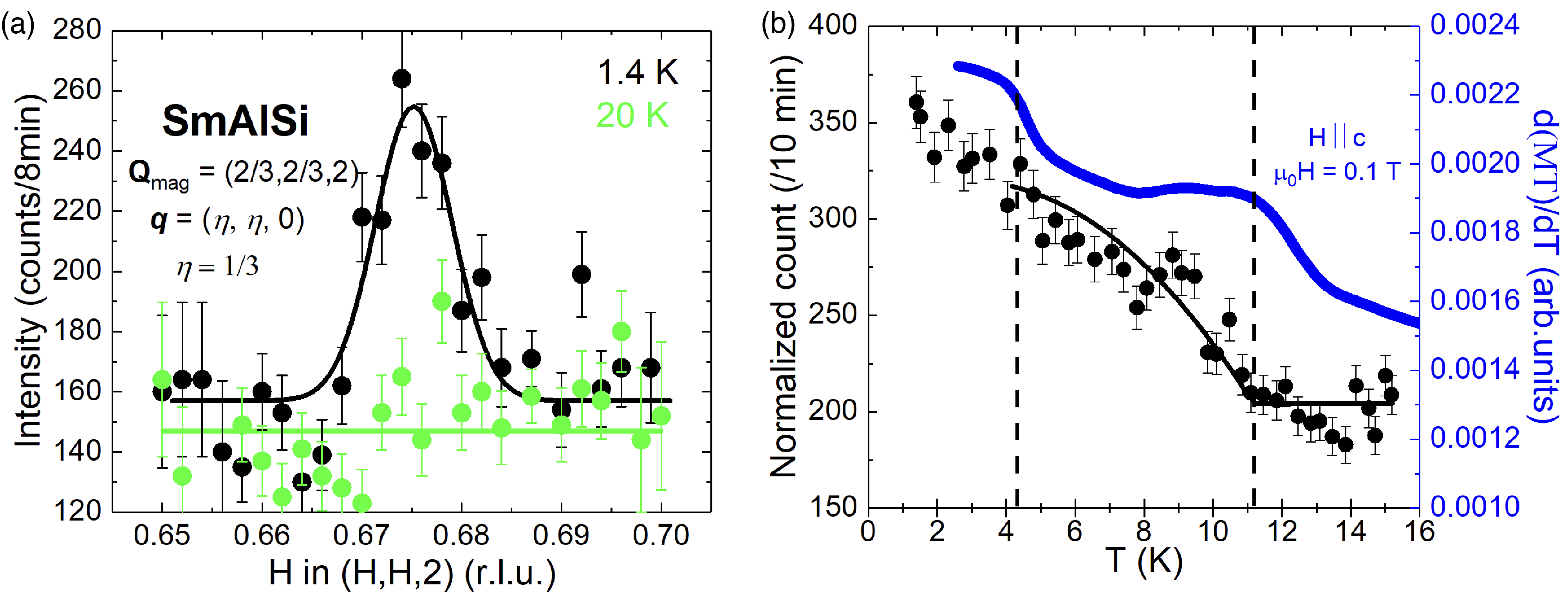} 
\caption{\label{SIfig:0} Elastic neutron scattering of SmAlSi. (a) Neutron diffraction scan along (H,H,2) direction. (b) Temperature dependence of the Bragg peak intensity (dots) and the mean field fit to the intensity (line). The blue line is d(MT)/dT for $H\parallel c$ and $\mu_0$H = 0.1 T. Error bars represent one standard deviation. }
\end{figure*}

The wave vector for the magnetic ordering has previously been determined to be close to (1/3, 1/3, 0), possibly slightly incommensurate with a value of (1/3-$\delta$, 1/3-$\delta$, 0) with $\delta$ = 0.007(7) \cite{Yao2022}.  We chose coarse collimation of 48´-40´-40´-120´ full-width-at-half-maximum to maximize the intensity, and consequently did not address the incommensurability issue. Initial exploratory measurements indicated that the signal-to-noise was somewhat better utilizing an energy of 14.7 meV; even though the absorption is higher, this was compensated by the higher Bragg peak structure factor. The quantitative determination of the magnetic structure(s) was prohibited by the very high Sm absorption.  Fig. \ref{SIfig:0}(a) shows the observed magnetic peak for the (2/3, 2/3, 2) reflection, which turned out to be strongest given the shape of the crystal. In Fig. \ref{SIfig:0}(b), the two magnetic transitions at T$_N$ = 11.3 K and $T\mathrm{_1}$ = 4.1 K are visible in the temperature dependence of the intensity of the (2/3, 2/3, 2) peak (circles, left axis) and the magnetization derivative d(MT)$/$dT (solid line, right axis).  

The black solid curve is a fit using mean field theory, from which the estimated ordering temperature is $T_N$ = 11.6(2) K.  We noted that the data appeared to have some extra intensity below the lower transition at $T\mathrm{_1}$ = 4.7 K, which may have a structural component but is currently not fully identified.  We therefore restricted the fit to temperatures above 5 K.  It is clear that mean field theory (or other typical order parameter models) does not adequately represent the observed temperature dependence of the data in the low temperature regime, which is quite different from the data in the inset to figure 2a in \cite{Yao2022}.  This indicates that there is additional magnetic intensity associated with this low temperature transition. The temperatures of the two transitions identified in neutron scattering measurements are consistent with our magnetization measurements (blue line in Fig. \ref{SIfig:0}(b)).

\section{Phase diagram determination}
\begin{figure*}[h]
\includegraphics[width=0.95\textwidth]{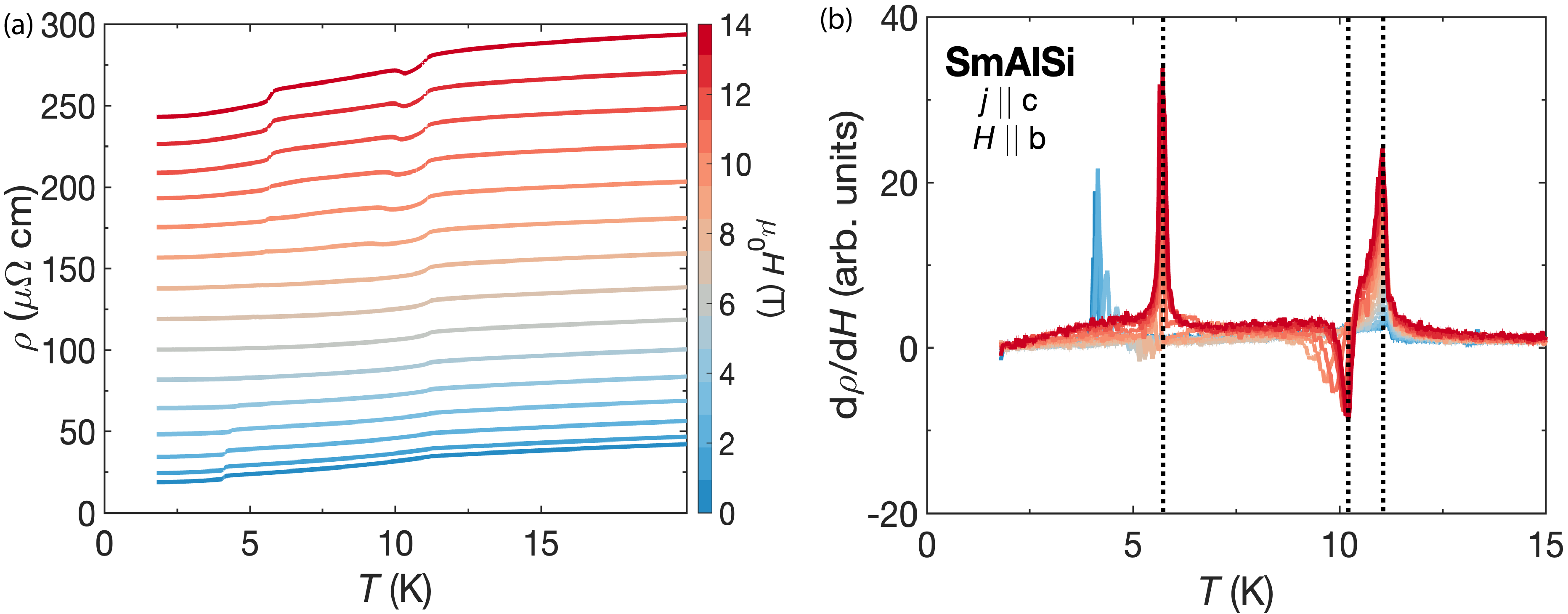} 
\caption{\label{SIfig:1} Temperature-dependent resistivity $\rho_{zz}$, with current $j\parallel c$ and magnetic field $H\parallel b$. The circles in Fig. 4(b) are determined from the maxima and minima in d$\rho$/d{\it T}.}
\end{figure*}

\begin{figure*}[h]
\includegraphics[width=0.95\textwidth]{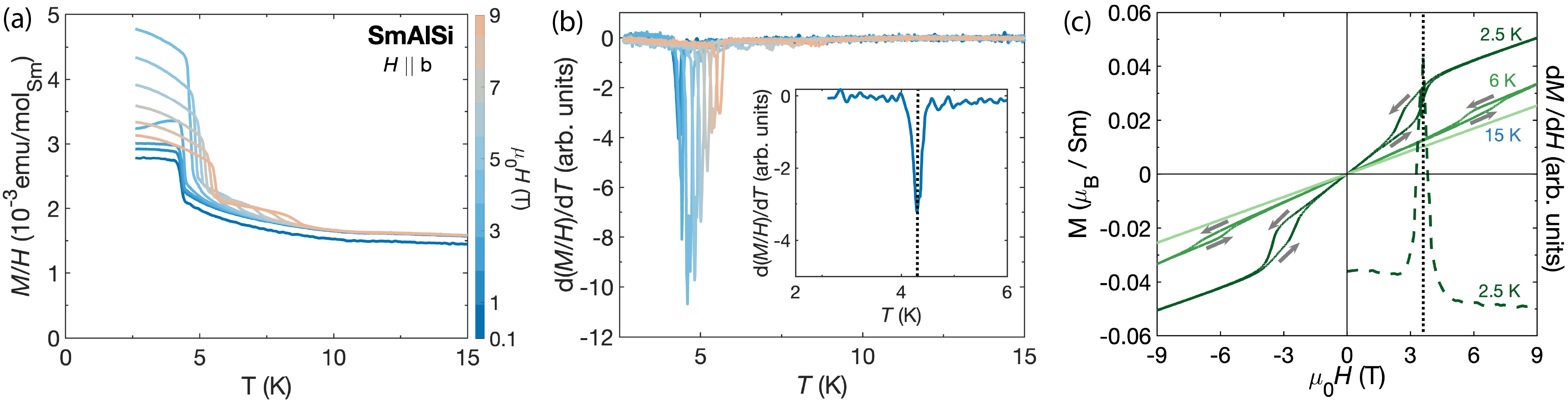}
\caption{\label{SIfig:2} (a) Temperature dependent susceptibility {\it M}/{\it H} with magnetic field $H\parallel b$. (b) Derivative of the M/H (a) as a function of temperature. (c) Magnetization as a function of field. The arrows indicate the field-swept direction. The dashed line is the derivative d{\it M}/d{\it H} at 2.5 K. The squares and diamonds in Fig. 4(a) are determined from the minima in d({\it M/H})/d{\it T} and the maxima in d{\it M}/d{\it H} as in (c), respectively.}
\end{figure*}

\begin{figure*}[h]
\includegraphics[width=0.6\textwidth]{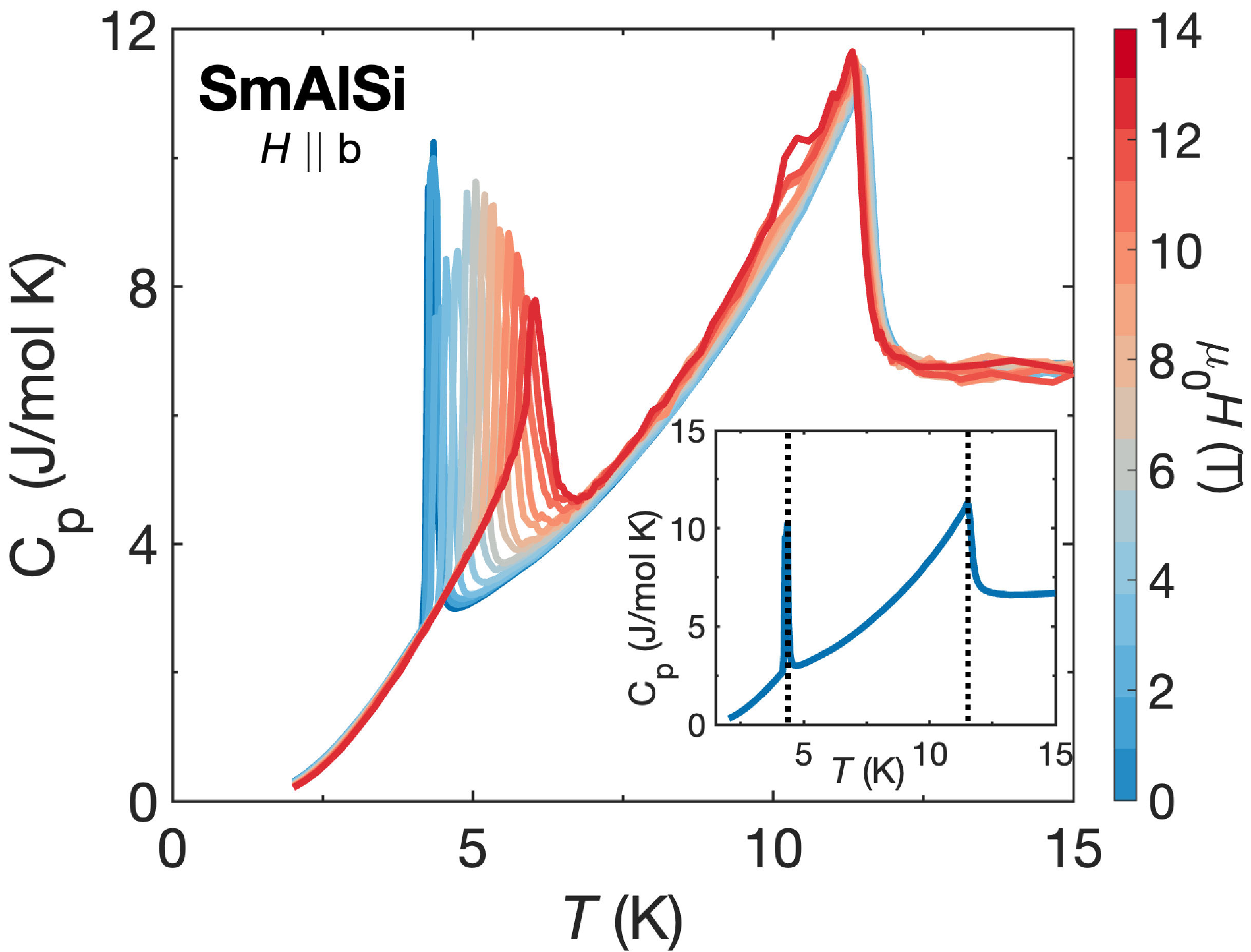} 
\caption{\label{SIfig:3} Temperature dependent heat capacity $C_p$ with magnetic field $H\parallel b$. The stars in Fig. 4(b) are determined from the peak positions.}
\end{figure*}

The magnetic phase diagram of SmAlSi is determined from electrical resistivity [Fig.~\ref{SIfig:1}], magnetization [Fig.~\ref{SIfig:2}] and heat capacity measurements [Fig.~\ref{SIfig:3}]. The phase boundaries in Fig. 4(b) are determined from the corresponding peaks in heat capacity [Fig. \ref{SIfig:3}] and the derivatives of resistivity and magnetization [Fig. \ref{SIfig:1}(b), Fig. \ref{SIfig:2}(b,c)].

\section{Hall effect}
The geometry for the Hall resistivity measurements $\rho_{xz}$ and $\rho_{zx}$ are shown in the inset of Fig. 2(a,b). According to the axes definition in Fig. 2(a,b), 
$\rho_{xz}=(V^+_x-V^-_x)*t/I$ and $\rho_{zx}=(V^+_z-V^-_z)*t/I$, 
where $V^{+/-}$ refers to the voltage, {\it I} is the current and {\it t} is the thickness of the sample. The data in the main text and supplementary materials are presented such that the positive (negative) sign of the signal reflects the positive (negative) sign of $V^+-V^-$ as defined in Fig. 2(a,b). Therefore, $-\sigma_{xz}$ and $-\rho_{zx}$ are shown instead of $\sigma_{xz}$ and $\rho_{zx}$.

\begin{figure*}[h]
\includegraphics[width=0.6\textwidth]{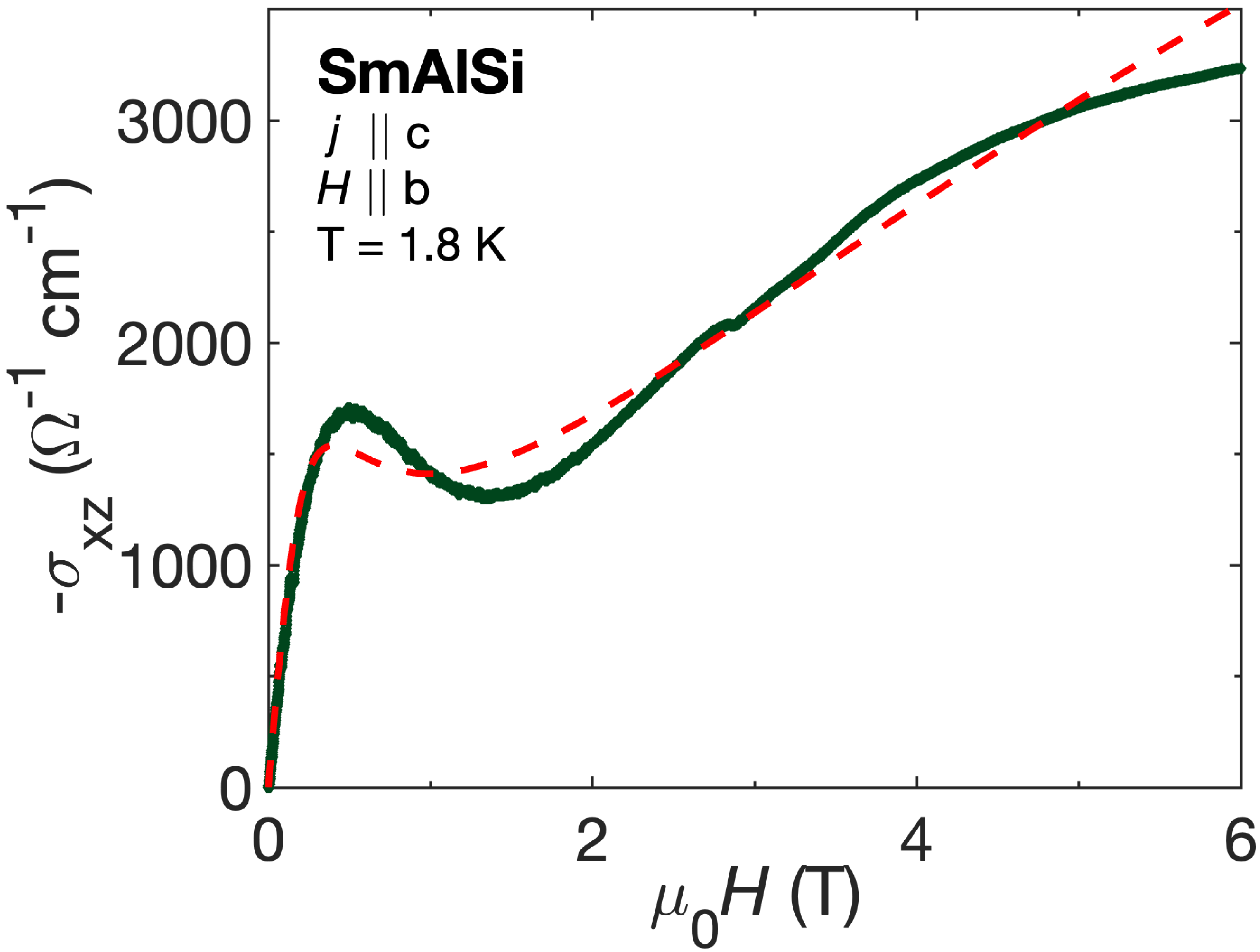} 
\caption{\label{SIfig:6} 
Two band model fitting to low field ($\mu_0$H $\leq$ 2.5 T) of $\sigma_{xz}$ at 1.8K with {\it n}$_1$ = 1.08*$\mathrm{10^{20}cm^{-3}}$, $\mu_1$ = 606$\mathrm{cm^{2} V^{-1}s^{-1}}$ and {\it n}$_2$ = 5.36*$\mathrm{10^{17}cm^{-3}}$, $\mu_2$ = 30436$\mathrm{cm^{2} V^{-1}s^{-1}}$.}
\end{figure*}

A two-band model fit attempting to capture the low field anomaly in $\sigma_{xz}$ is shown in Fig.~\ref{SIfig:6}. As discussed in the main text, the fit does not capture the Hall conductivity well, and the resulting longitudinal conductivity $\sigma_{xx}$ is significantly lower than the $\sigma_{xx}$ from measurements, which suggests that a simple multi-band model is not an adequate description of the Hall conductivity. 

\begin{figure*}[h]
\includegraphics[width=0.95\textwidth]{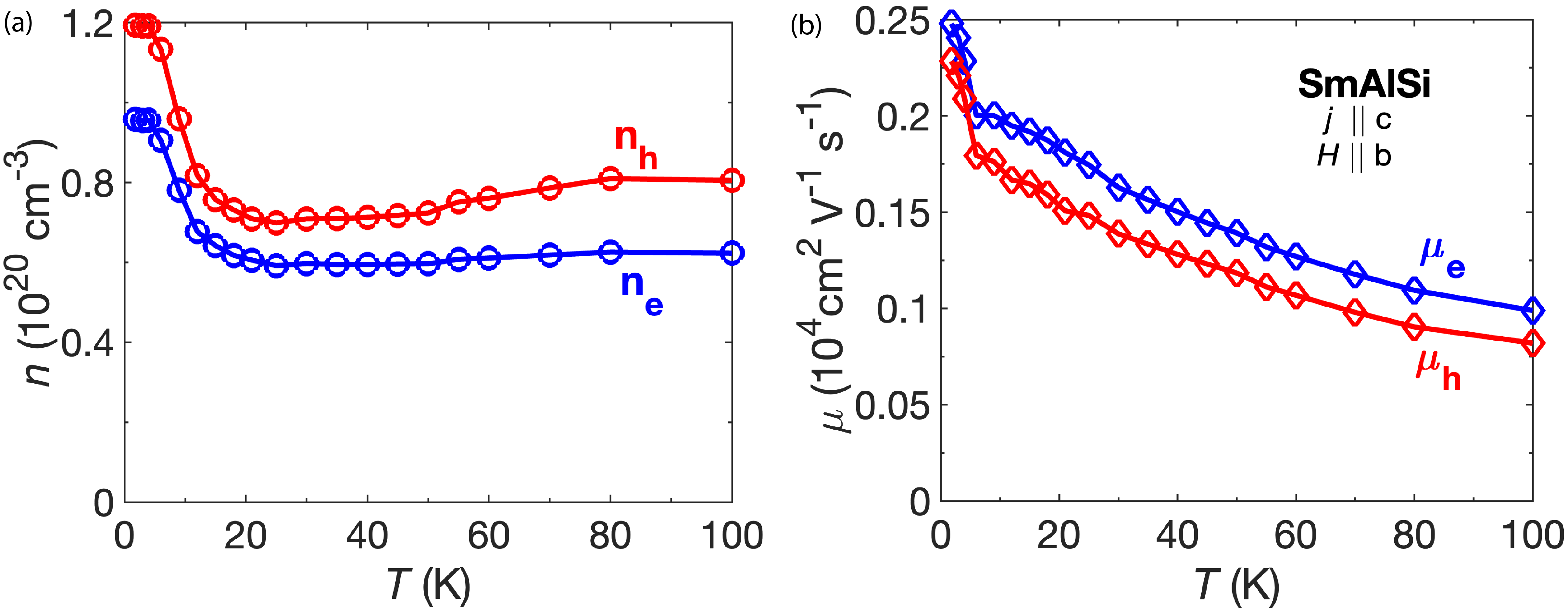} 
\caption{\label{SIfig:7}
The carrier concentration $n_e$, $n_h$ and carrier mobility $\mu_e$, $\mu_h$ extracted from the two-band model fitting on the Hall conductivity $\sigma_{xz}$ to high fields ($\mu_0$H $\geq$ 2.5 T) as a function of temperature.}
\end{figure*}

The normal Hall effect is extracted through a two-band model fitting to the high field range ($\mu_0${\it H} $\geq$ 2.5 T). These fits reveal the coexistence of a hole pocket and an electron pocket. The carrier concentration and carrier mobility from the fits are shown in Fig. \ref{SIfig:7}. Both the carriers concentrations and mobility are nearly temperature-independent, implying that the Fermi surface of SmAlSi does not undergo reconstruction.




\section{Quantum oscillations}

\begin{figure*}
\includegraphics[width=0.95\textwidth]{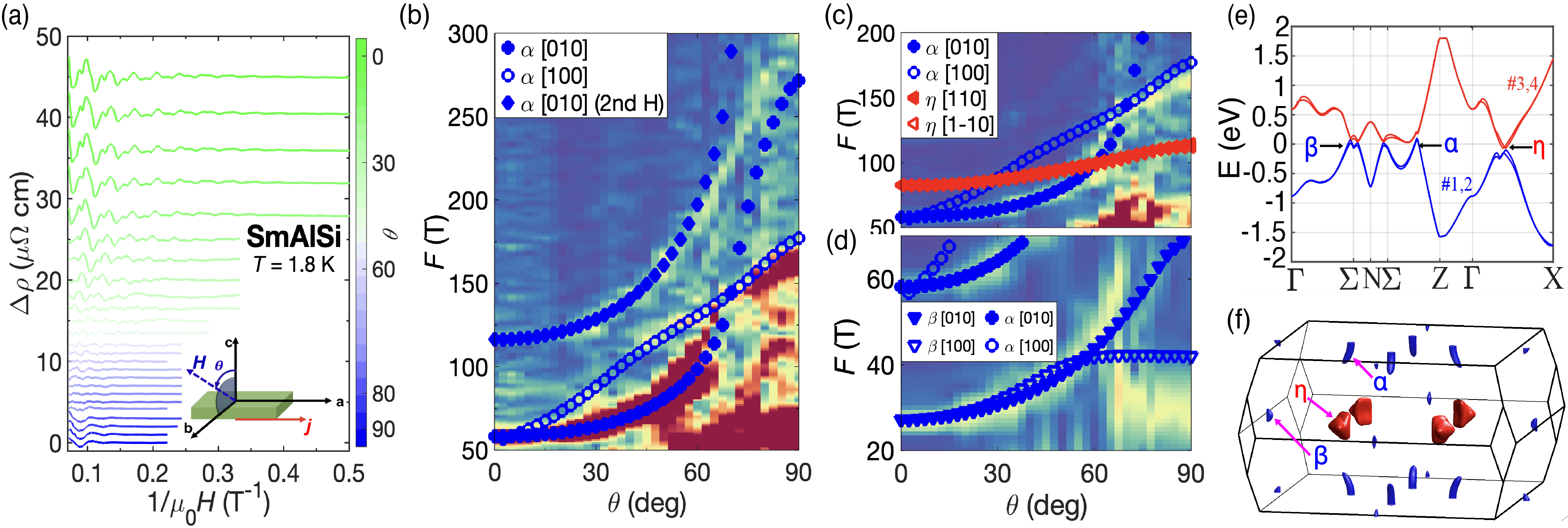} 
\caption{\label{SIfig:11} Angle-dependent SdH oscillations of SmAlSi for {\it T} = 1.8 K (ordered state). (a) SdH oscillations from angle-dependent resistivity measurements after background subtraction. The measurement geometry is plotted in the inset. (b-d) Contour map of the FFT spectra in (a), with an emphasis on high frequency (b), medium frequency (c), and low frequency (d). The symbols represent calculated frequencies from cross-sections in DFT calculations. The [100] and [010] refers to the pockets that are located at $k_x$ and $k_y$ axis, respectively. (e) Partial band structure of SmAlSi in the PM state from DFT calculations \cite{Zhang2022}. Only the bands that cross the Fermi energy are shown here. (f) Partial Fermi surface from (e) showing the location of the Fermi pockets identified in Fig. \ref{SIfig:11}(b-d).}
\end{figure*}



\begin{figure*}[h]
\includegraphics[width=0.95\textwidth]{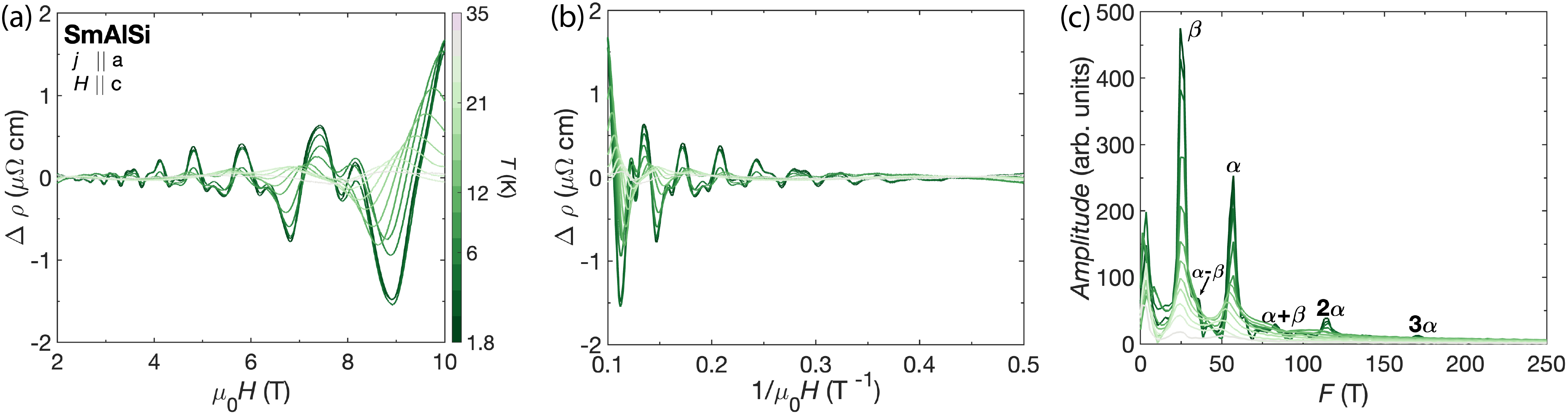} 
\caption{\label{SIfig:10} 
Temperature-dependent SdH oscillations of SmAlSi. (a,b) Resistivity after background subtraction plotted as a function of the magnetic field $\mu_0H$ (a) and a function of 1/$\mu_0H$ (b). (c)FFT spectra of the results in (b). }
\end{figure*}

Shubnikov-de Haas (SdH) oscillations are used to understand the Fermi surface and how it changes with temperature. The oscillation frequency \textit{F} from QO measurements is related to the extremal cross-sectional area $A_k$ of the FS perpendicular to the applied magnetic field via the Onsager relation: $F=(\hbar A_k)/(2\pi e)$, where $\hbar$ is the reduced Planck constant and \textit{e} is the electron charge. The angular evolution of the QO frequency reflects the change of the extreme orbits as the sample rotates relative to the magnetic field direction, while the change in QO frequency with temperature indicates a change in the extremal FS cross section, and might indicate a non-trivial FS topology. For both sets of measurements, the current is applied along the \textit{a} direction, and the field is kept within the \textit{bc} plane during the rotation for angle-dependent measurements. An illustration of the angle-dependent measurement configuration is shown in the inset of Fig. \ref{SIfig:11}(a).

We measured the SdH oscillations at different temperatures, and the background-subtracted results are shown in Fig.~\ref{SIfig:10}(a,b). The oscillation amplitude is lower at higher temperatures, as expected. The frequency peaks in the FFT spectra are labeled according to the definition in Fig.~\ref{SIfig:11} (e,f) \cite{Zhang2022}. The QO frequencies do not change across the magnetic transition temperatures $T\mathrm{_N}$ and $T\mathrm{_1}$. Instead, a gradual change in the frequencies in the FFT spectrum is found above 15 K, which is higher than $T\mathrm{_N}$. This is uncommon since the FS reconstruction is not expected in the paramagnetic state.

It should be noted that a similar shift has been found in previous SmAlSi  studies \cite{Cao2022}, and also in the isostructural NdAlSi \cite{Gaudet2021}. In Fig. 4(b, c) of \cite{Gaudet2021} it can be seen that the QO frequencies are also shifting within the same magnetic phases at different temperatures in NdAlSi.

Moreover, Wang et al \cite{Wang2023} discovered QOs as a function of temperature at a constant magnetic field in the PM state, where the exchange interaction between the conduction electrons and 4f elections appeared to be crucial. They propose that these signatures are due to the destructive interference between two QOs originating from the FS splitting, accompanied by a strong temperature- and magnetic field-dependent phase inversion.

Such QOs as a function of temperature are not discovered in SmAlSi. The exchange interaction between the conduction electrons and 4f elections is expected to be significantly smaller in SmAlSi, given the small size of the $Sm^{3+}$ effective moment ($\mu_{eff}$ = 0.714 $\mu_{B}$) and polarized moment under field (M = 0.05 $\mu_{B}$/$Sm^{3+}$  at {\it T} = 1.8 K and $\mu_{0}${\it H} = 9 T). Therefore, the Fermi surface splitting in SmAlSi is dramatically suppressed.

We want to point out another possible reason for the quantum oscillation frequency shift. As shown in \cite{Guo2021}, the thermal broadening of the Fermi energy due to finite temperature will induce a frequency decrease with increasing temperature. Such an effect depends on the effective mass of the quasiparticles and is enhanced if the band dispersion is linear. The derivation in \cite{Guo2021} is carried out in magnetic ions, which may further enhance the effect. This may explain the frequency shift of the frequency $\alpha$.

The physics behind the QO frequency shift requires further investigation. This is outside the focus of this work and we will leave it to future studies.

The angle-dependent measurements are obtained at 1.8 K in the ordered state. We also performed the same measurements at 12 K in the PM state for completeness and to compare with results in \cite{Zhang2022}. The resulting SdH oscillations at various angles $\theta$ with $0~\leq \theta \leq 90^\circ$ are shown in Fig. \ref{SIfig:11}(a). Similar to the QOs at 12 K, QOs are observed in the entire range of $\theta$. For T = 1.8 K, QOs onset can be observed starting at relatively low fields ($\sim$ 3 T), indicative of high crystal quality. 

To build a correspondence between QOs and the FS, a comparison between experiments and calculation is required. The electronic structure of a magnetic material is correlated with the magnetic order. However, the large neutron absorption cross-section of the Sm atoms and the incommensurate structure make it difficult to resolve the magnetic structure from neutron scattering. Our isothermal magnetization measurements {\it M(H)} [shown in Fig. \ref{SIfig:2}(c)] indicate that the net magnetization is smaller than 0.05 $\mu_B$/Sm$^{3+}$ up to the highest measured field of 9~T. Therefore, the magnetic order under field remains dominantly AFM and possibly retains the incommensurate nature. Given the uncertainty in determining the magnetic structure and the limitations of band structure calculations with incommensurate magnetic configuration, we compare the experimental QO results with the DFT prediction in the PM state. That is validated by: (1) the fact that the QO frequency is unchanged across the transition temperature and (2) the small moment of Sm$^{3+}$ ($\mu_{sat}^{Sm^{3+}}~=~0.714~\mu_B$) 3. In a recent ARPES study by by Lou \textit{et al.} \cite{Lou2023}, no evident evolution of the band structure across the magnetic transitions . 

Therefore, we used our calculations (reported in \cite{Zhang2022}) to match the experimental QO frequencies. The energy bands that cross the Fermi energy are shown in Fig.~\ref{SIfig:11}(e). At a first glance, four bands appear to cross the Fermi energy, forming multiple Fermi pockets \cite{Zhang2022}. However, the four bands that form these pockets are singly-degenerate, where the degeneracy is lifted by SOC and broken inversion symmetry \textit{I}. We argue that consideration of only the spin-majority band is adequate here, for the following two reasons: the energy dispersion of the spin-majority/minority bands and the shape of the Fermi pockets from these bands is similar; under an external magnetic field, the conduction electrons are partially polarized along the field orientation, and the QOs of the corresponding bands are expected to dominate. Upon careful mapping between the extremal cross sections in theory and QO frequencies in experiments, we can identify the underlying Fermi pockets that contribute to the QOs, as shown in Fig. \ref{SIfig:11}(b-d). The contour map of the QO frequencies at 1.8 K is similar to that at 12 K. The contour map of the QO frequencies consists of the following contributions: fundamental oscillations of pocket $\alpha$ (blue circles), the second harmonic of pocket $\alpha$ (blue diamonds), fundamental oscillations of pocket $\beta$ (blue triangles), and fundamental oscillations of pocket $\eta$ (red triangles). These Fermi pockets are plotted in Fig.~\ref{SIfig:11}(f). Compared to our previous results at 12 K \cite{Zhang2022}, in addition to pocket $\alpha$ and pocket $\beta$, we found frequencies corresponding to pocket $\eta$ and second harmonic of pocket $\alpha$. The similarity of the contour maps between 1.8 K and 12 K also implies that it is unlikely to undergo a Fermi surface reconstruction in SmAlSi. These Fermi pockets are relevant to the Weyl points W1 and W2 that are close to the Fermi energy \cite{Zhang2022}.